%% file: main.tex
\documentclass[twocolumn,trackchanges]{aastex63}
\pdfoutput=1
\usepackage[T1]{fontenc}
\usepackage{fontawesome}
\usepackage{color}
\usepackage{amsmath}
\usepackage{natbib}
\usepackage{ctable}
\usepackage{bm}
\usepackage[normalem]{ulem} 
\usepackage{xspace}
\usepackage{paralist}
\usepackage{fontawesome}

\defcitealias{zheng2007}{Z07}
\defcitealias{simbig_letter}{H22a}
\defcitealias{simbig_mock}{H23}

\newcommand{\given}{\,|\,}

\newcommand{\bfi}[1]{\textbf{\textit{#1}}}

\newcommand{\eg}{\emph{e.g.}}
\newcommand{\ie}{\emph{i.e.}}

\let\oldAA\AA
\renewcommand{\AA}{\text{\normalfont\oldAA}}

\newcommand{\btheta}{\boldsymbol{\theta}}
\newcommand{\bphi}{\boldsymbol{\phi}}

\newcommand{\simbig}{{\sc SimBIG}}

\newcommand{\bitem}{\begin{itemize}}
\newcommand{\eitem}{\end{itemize}}
\newcommand{\beq}{\begin{equation}}
\newcommand{\eeq}{\end{equation}}

\definecolor{orange}{rgb}{1,0.5,0}

\begin{document} 

\title{\simbig: The First Cosmological Constraints from the Non-Linear Galaxy Bispectrum}

\author[0000-0003-1197-0902]{ChangHoon Hahn}
\altaffiliation{changhoon.hahn@princeton.edu.com}
\affil{Department of Astrophysical Sciences, Princeton University, Princeton NJ 08544, USA} 

\author{Michael Eickenberg}
\affil{Center for Computational Mathematics, Flatiron Institute, 162 5th Avenue, New York, NY 10010, USA}

\author{Shirley Ho}
\affil{Center for Computational Astrophysics, Flatiron Institute, 162 5th Avenue, New York, NY 10010, USA}

\author{Jiamin Hou}
\affil{Department of Astronomy, University of Florida, 211 Bryant Space Science Center, Gainesville, FL 32611, USA}
\affil{Max-Planck-Institut f\"ur Extraterrestrische Physik, Postfach 1312, Giessenbachstrasse 1, 85748 Garching bei M\"unchen, Germany}

\author{Pablo Lemos}
\affil{Department of Physics, Universit\'{e} de Montr\'{e}al, Montr\'{e}al, 1375 Avenue Th\'{e}r\`{e}se-Lavoie-Roux, QC H2V 0B3, Canada}
\affil{Mila - Quebec Artificial Intelligence Institute, Montr\'{e}al, 6666 Rue Saint-Urbain, QC H2S 3H1, Canada}
\affil{Center for Computational Mathematics, Flatiron Institute, 162 5th Avenue, New York, NY 10010, USA}

\author[0000-0002-0637-8042]{Elena Massara}
\affil{Waterloo Centre for Astrophysics, University of Waterloo, 200 University Ave W, Waterloo, ON N2L 3G1, Canada}
\affil{Department of Physics and Astronomy, University of Waterloo, 200 University Ave W, Waterloo, ON N2L 3G1, Canada}

\author{Chirag Modi}
\affil{Center for Computational Mathematics, Flatiron Institute, 162 5th Avenue, New York, NY 10010, USA}
\affil{Center for Computational Astrophysics, Flatiron Institute, 162 5th Avenue, New York, NY 10010, USA}

\author[0000-0001-8841-9989]{Azadeh Moradinezhad Dizgah}
\affil{D\'epartement de Physique Th\'eorique, Universit\'e de Gen\`eve, 24 quai Ernest Ansermet, 1211 Gen\`eve 4, Switzerland}

\author{Liam Parker}
\affil{Department of Astrophysical Sciences, Princeton University, Princeton NJ 08544, USA} 

\author[0000-0003-0055-0953]{Bruno R\'egaldo-Saint Blancard}
\affil{Center for Computational Mathematics, Flatiron Institute, 162 5th Avenue, New York, NY 10010, USA}

\begin{abstract}
    We present the first cosmological constraints from analyzing higher-order galaxy clustering on non-linear scales. 
    We use \simbig, a forward modeling framework for galaxy clustering analyses that employs 
    simulation-based inference to perform highly efficient cosmological inference using normalizing flows.  
    It leverages the predictive power of high-fidelity simulations and robustly
    extracts cosmological 
    information from regimes inaccessible with current standard analyses. 
    In this work, we apply \simbig~to a subset of the BOSS galaxy sample and analyze 
    the redshift-space bispectrum monopole, $B_0(k_1, k_2, k_3)$, to  $k_{\rm max}=0.5\,h/{\rm Mpc}$. 
    We achieve 1$\sigma$ constraints of $\Omega_m=0.293^{+0.027}_{-0.027}$ and 
    $\sigma_8= 0.783^{+0.040}_{-0.038}$, which are more than 1.2 and 2.4$\times$ tighter 
    than constraints from standard power spectrum analyses of the same dataset. 
    We also derive 1.4, 1.4, 1.7$\times$ tighter constraints on $\Omega_b$, $h$, $n_s$.
    This improvement comes from additional cosmological information in 
    higher-order clustering on non-linear scales and, for $\sigma_8$, is 
    equivalent to the gain expected from a standard analysis on a $\sim$4$\times$ larger galaxy sample. 
    Even with our BOSS subsample, which only spans 10\% of the full BOSS volume, 
    we derive competitive constraints on the growth of structure: 
    $S_8 = 0.774^{+0.056}_{-0.053}$.
    Our constraint is consistent with results from both cosmic microwave background and weak lensing. 
    Combined with a $\omega_b$ prior from Big Bang Nucleosynthesis, we also derive a
    constraint on $H_0=67.6^{+2.2}_{-1.8}\,{\rm km\,s^{-1}\,Mpc^{-1}}$ that is consistent with 
    early universe constraints.
\end{abstract} 
\keywords{cosmological parameters from LSS --- Machine learning --- cosmological simulations --- galaxy surveys}

\input{intro}
\input{obs}
\input{fm}
\input{results}
\input{discuss}
\input{summary}

\section*{Acknowledgements}
It's a pleasure to thank 
Mikhail M. Ivanov and Yosuke Kobayashi for providing us with the posteriors used for comparison. 
We also thank Peter Melchior, Uro{\u s}~Seljak, and Benjamin D. Wandelt for valuable discussions. 
This work was supported by the AI Accelerator program of the Schmidt Futures Foundation. JH has received funding from the European Union's Horizon 2020 research and innovation program under the Marie Sk\l{}odowska-Curie grant agreement No 101025187. AMD acknowledges funding from Tomalla Foundation for Research in Gravity. 

\bibliography{simbig} 
\end{document}

%% file: intro.tex
\section{Introduction} \label{sec:intro} 
The three-dimensional spatial distribution of galaxies enables us to 
constrain the nature of dark matter and dark energy and measure the 
contents of the Universe. 
Along with other cosmological probes, it provides one of the most stringent 
tests of the standard $\Lambda$CDM cosmological model that can lead to 
discoveries of new physics.
With this aim, spectroscopic galaxy surveys of the next decade, the 
Dark Energy Spectroscopic Instrument~\citep[DESI;][]{desicollaboration2016, desicollaboration2016a, abareshi2022}, 
Subaru Prime Focus Spectrograph~\citep[PFS;][]{takada2014, tamura2016}, 
the ESA {\em Euclid} satellite mission~\citep{laureijs2011}, and the
Nancy Grace Roman Space Telescope~\citep[Roman;][]{spergel2015, wang2022a}, 
will probe galaxies over unprecedented cosmic volumes out to $z\sim3$.  

Current analyses of galaxy clustering focus on the power spectrum,
the Fourier counterpart to the two-point correlation function, as the primary 
measurement of galaxy clustering~\citep[\emph{e.g.}][]{beutler2017, ivanov2020, chen2022, kobayashi2022}. 
These standard analyses model the power spectrum using the perturbation theory 
(PT) of large-scale structure~\citep[see][for a review]{bernardeau2002,desjacques2016}. 
As a result, they focus on large, mostly linear, scales 
($k_{\rm max} \sim 0.2\,h/{\rm Mpc})$ where deviations from linear theory are 
small and PT remains valid. 
Accurate modeling of higher-order clustering statistics (\emph{e.g.} bispectrum) 
with PT is progressively more complex and challenging. 
Furthermore, there are currently no PT-based models that describe new promising 
summary statistics~\citep[\emph{e.g.}][]{banerjee2021, eickenberg2022, valogiannis2022, naidoo2022}. 

Meanwhile, recent studies have now established that there is additional 
cosmological information in higher-order 
statistics~\citep[\emph{e.g.}][]{gil-marin2017, damico2022, philcox2022}. 
Forecasts have also long suggested that there may be even more information 
on small scales~\citep[\emph{e.g.}][]{sefusatti2005}.
Recently, \cite{hahn2020} and \cite{hahn2021a} showed that constraints 
on $\Lambda$CDM cosmological parameters, 
$\Omega_m, \Omega_b, h, n_s, \sigma_8$, improve by a factor of
$\sim$2 by analyzing the bispectrum down to non-linear scales 
($k_{\rm max} =  0.5\,h/{\rm Mpc}$).
\cite{massara2020, gualdi2021, massara2022, wang2022, hou2022, eickenberg2022, valogiannis2022, porth2023} 
found consistent improvements from forecasts of other summary statistics that 
extract non-Gaussian cosmological information from non-linear
scales. 
These improvements are further corroborated by recent small-scale clustering analyses 
using emulators~\citep{storeyfisher2022, zhai2022}. 

Another major limitation of current analyses is robustly accounting for observational 
systematics in \emph{e.g.} targeting, imaging, completeness that significantly impact 
clustering measurements~\citep[][]{ross2012, ross2017}.
Fiber collisions, for example, prevent galaxy surveys that use fiber-fed 
spectrographs (\emph{e.g.} DESI, PFS) from successfully measuring redshifts from galaxies 
within some angular scale of one another~\citep{yoon2008}.
They significantly bias the power spectrum measurement on scales smaller than 
$k > 0.1\,h/{\rm Mpc}$~\citep{guo2012, hahn2017a, bianchi2018}.
While improved correction schemes for fiber collisions may be sufficient for power
spectrum analyses~\citep{hahn2017a, pinol2017, bianchi2018, smith2019},
no correction scheme has yet been designed or demonstrated for other summary statistics. 

Recently, \cite{simbig_letter} and \cite{simbig_mock}\footnote{
hereafter \citetalias{simbig_letter} and \citetalias{simbig_mock}}
presented the SIMulation-Based 
Inference of Galaxies (\simbig), a forward modeling framework for analyzing galaxy 
clustering. 
\simbig~uses simulation-based 
inference\footnote{also known as ``likelihood-free inference'' (LFI) or ``implicit likelihood inference'' (ILI)}~\citep[SBI; see][for a review]{cranmer2020}
to perform highly efficient cosmological parameter inference using neural density estimation (NDE)
from machine learning~\citep[\emph{e.g.}][]{germain2015, papamakarios2017}.
This enables \simbig~to use high-fidelity simulations that model the details and 
realism of the observations. 
In particular, the \simbig~forward model is based on cosmological $N$-body simulations 
that can more accurately model non-linear structure formation to smaller scales than PT. 
It also includes observational systematics 
(\emph{e.g.} survey geometry, masking, fiber collisions).
With this approach, \citetalias{simbig_letter} analyzed the galaxy power spectrum from the 
Sloan Digital Sky Survey (SDSS)-III Baryon Oscillation Spectroscopic 
Survey~\citep[BOSS;][]{eisenstein2011, dawson2013}.
This work demonstrated that they can rigorously analyze the power spectrum down to
smaller scales than ever before, $k_{\rm max} = 0.5\,h/{\rm Mpc}$.

In this work, we extend the \simbig~analysis to the first higher-order statistic: the bispectrum.
For a near-Gaussian galaxy distribution, the bispectrum extracts nearly all of its 
cosmological information~\citep[\emph{e.g.}][]{fry1994, matarrese1997, scoccimarro2000}.
We present the first robust cosmological constraints from an analysis that exploits
clustering information on both non-linear scales and in higher-order statistics.
We begin in Section~\ref{sec:obs} by describing the observational galaxy sample
that we analyze.
We then briefly summarize the details of the \simbig~approach in Section~\ref{sec:simbig}.
We present and discuss our cosmological results in Section~\ref{sec:results} and compare
them to constraints in the literature. 

%% file: obs.tex
\section{Observations: BOSS CMASS Galaxies} \label{sec:obs}
We apply our \simbig~bispectrum analysis to the same observed galaxy sample as 
\citetalias{simbig_letter}, which is derived from the Sloan Digital Sky Survey 
(SDSS)-III Baryon Oscillation Spectroscopic Survey (BOSS) Data Release 
12~\citep{eisenstein2011, dawson2013}.
More specifically, the sample consists of galaxies in the Southern
Galactic Cap (SGC) of BOSS CMASS galaxy sample that are within the redshift
range $0.45 < z < 0.6$ and have ${\rm Dec} > -6$ deg. and $-25 < {\rm RA} < 28$ deg.
Overall, the galaxy sample covers $\sim$3,600 ${\rm deg}^{2}$ and includes
109,636 galaxies. 
This corresponds to 70\% of the SGC footprint and $\sim$10\% of the 
full BOSS volume. 
We refer readers to \citetalias{simbig_letter} and \citetalias{simbig_mock} for further 
details on the observed galaxy sample. 

%% file: fm.tex
\section{\simbig~with the Galaxy Bispectrum} \label{sec:simbig}

The {\sc SimBIG} approach uses SBI to infer posteriors of $\Lambda$CDM 
cosmological parameters with only a forward model that can 
generate mock observations, 
\emph{i.e.} the 3D galaxy distribution. 
In this section, we briefly describe the forward model, the SBI 
methodology, the bispectrum, and our posterior validation.

\subsection{Forward Model} \label{sec:fm} 
The {\sc Simbig} forward model constructs simulated galaxy catalogs
from {\sc Quijote} $N$-body simulations run at different cosmologies 
in a Latin-hypercube configuration~\citep{villaescusa-navarro2020}.  
Each simulation has a volume of $1\,(h^{-1}{\rm Gpc})^3$ and is
constructed using $1024^3$ cold dark matter (CDM) particles 
gravitationally evolved from $z=127$ to $z=0.5$.
From the $N$-body simulations, halos are identified using the 
phase-space information of dark matter particles with the 
{\sc Rockstar} halo finder~\citep{behroozi2013a}.
Afterwards, the halos are populated using the halo occupation 
distribution~\citep[HOD;][]{berlind2002, zheng2007} framework, which 
provides a flexible statistical prescription for determining the 
number of galaxies as well as their positions and velocities within 
halos.  
{\sc SimBIG} uses a state-of-the-art HOD model with 9 parameters 
that supplements the standard~\cite{zheng2007} model 
with assembly, concentration,  and velocity biases. 

From the HOD galaxy catalog, {\sc SimBIG} adds a full BOSS survey 
realism by applying the exact survey geometry and observational 
systematics. 
The forward modeled catalogs have the same redshift range and angular 
footprint of the CMASS sample, including masking for bright stars,
centerpost, bad field, and collision priority. 
Furthermore, {\sc SimBIG} also includes fiber collisions, which 
systematically removes galaxies in galaxy pairs within an angular
scale of $62^{\prime\prime}$. 
We forward model fiber collisions because the standard correction schemes 
do not accurately correct for them~\citep{hahn2017a}.  
In summary, the \simbig~forward model aims to generate mock 
galaxy catalogs that are statistically indistinguishable from the 
observations.
For more details on the forward model, we refer readers to 
\citetalias{simbig_letter} and \citetalias{simbig_mock}.

\begin{figure*}
\begin{center}
    \includegraphics[width=\textwidth]{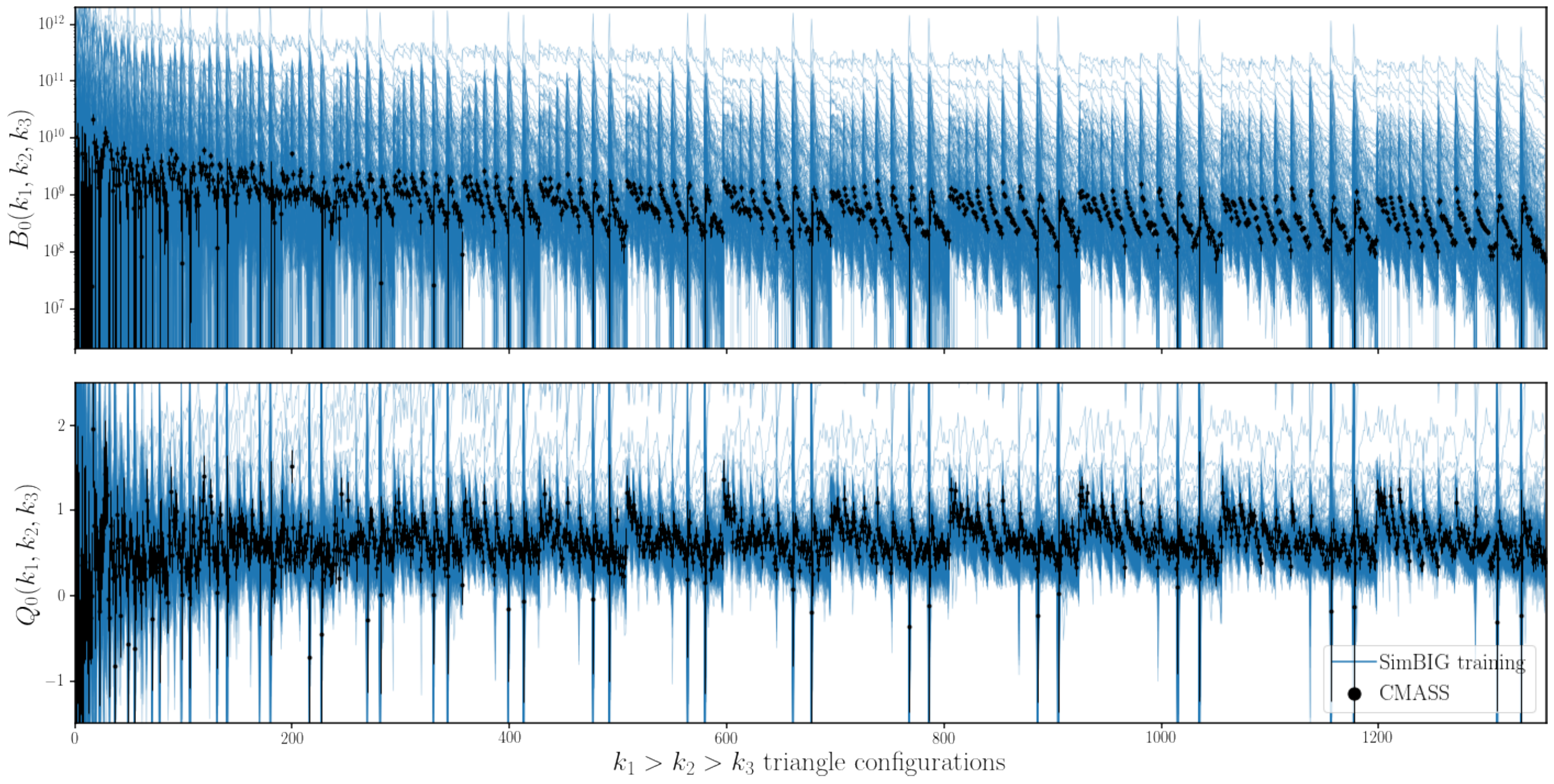}
    \caption{
        The bispectrum monopole ($B_0$; top panel) and reduced bispectrum
        monopole ($Q_0$; bottom panel) of a subset of simulated galaxy
        catalogs in our training set. 
        The catalogs are constructed using the \simbig~forward model from 
        the {\sc Quijote} $N$-body simulations and include BOSS survey realism.
        We randomly select 200 out of the 20,000 catalogs.
        We present a subset of 1,354 triangle configurations with 
        $k_1, k_2, k_3 < k_{\rm max} = 0.25\,h/{\rm Mpc}$,for clarity. 
        The configurations are ordered by looping through $k_3$ in the inner most 
        loop and $k_1$ in the outer most loop with $k_1 \leq k_2 \leq k_3$.
        For reference, we include $B_0$ measured from the observed 
        BOSS CMASS sample (black) with errorbars estimated from the 
        $\mathtt{TEST0}$ simulations.
        The observed $B_0$ is well within our training dataset.
    }\label{fig:b0k}
\end{center}
\end{figure*}

\subsection{Simulation-Based Inference} \label{sec:sbi} 
From the forward modeled galaxy catalogs, we use the {\sc SimBIG} 
SBI framework to infer posterior distributions of cosmological 
parameters, $\btheta$, for a given summary statistic, $\bfi{x}$,
of the observations: $p(\btheta\given\bfi{x})$. 
The {\sc SimBIG} SBI framework enables cosmological inference with a limited 
number of forward modeled simulations. 
This in turn enables us to exploit cosmological information on small,
non-linear, scales and in higher-order statistics that is inaccessible
with standard cosmological analyses. 

The SBI in \simbig~is based on NDE and uses ``normalizing flow''
models~\citep{tabak2010, tabak2013, rezende2015}. 
Normalizing flows use neural networks to learn an extremely flexible 
and bijective transformation, $f: x \mapsto z$, that maps a complex target
distribution to a simple base distribution, $\pi(\bfi{z})$, that is fast to evaluate.
$f$ is defined to be invertible and have a tractable Jacobian so that the 
target distribution can be evaluated from $\pi(\bfi{z})$ by change of 
variables. 
Since $\pi(\bfi{z})$ is easy to evaluate, this enables us to also easily 
evaluate the target distribution.
In our case, the target distribution is the posterior and the base distribution 
is a multivariate Gaussian. 
Among various normalizing flow architectures, we use Masked Autoregressive
Flow~\citep[MAF;][]{papamakarios2017} models.\footnote{We use the MAF 
implementation in $\mathtt{sbi}$ Python package~\citep{greenberg2019,
tejero2020}, 
which is based on the $\mathtt{nflows}$ Python
package~\citep{durkan2019_nflows, nflows}.
}. 

Our goal is to train a normalizing flow with hyperparameters, $\bphi$, 
that best approximates the posterior, 
$q_{\bphi}(\btheta\given\bfi{x}) \approx p(\btheta\given\bfi{x})$. 
We do this by minimizing the forward KL divergence between 
$p(\btheta, \bfi{x}) = p(\btheta\given\bfi{x}) p(\bfi{x})$ and
$q_{\bphi}(\btheta\given\bfi{x}) p(\bfi{x})$. 
In practice, we first split the forward modeled catalogs into a 
training and validation set with a 90/10 split.
Then we maximize the total log-likelihood 
$\sum_i \log q_{\bphi}(\btheta_i\given \bfi{x}_i)$ over the 
training set, $\{(\btheta_i, \bfi{x}_i)\}$
This is equivalent to minizmizing the forward KL divergence.
We use the {\sc Adam} optimizer~\citep{kingma2017} with a batch size of 50. 
To prevent overfitting, we evaluate the total log-likelihood on the validation
data at every training epoch and stop the training when the validation 
log-likelihood fails to increase after 20 epochs.  

We determine the architecture of our normalizing flow, \ie~number of blocks,
transforms, hidden features, and dropout probability, through experimentation.
We train a large number of flows with architectures and learning rates determined 
using the {\sc Optuna} hyperparameter optimization framework~\citep{akiba2019}.
Afterwards, we select five normalizing flows with the lowest validation losses. 
Our final flow is an equally weighted ensemble of the flows: 
$q_{\bphi}(\btheta\given\bfi{x})  =  \sum_{j=1}^5 q_\phi^j(\btheta\given\bfi{x})/5$. 
We find that ensembling flows with different initializations and 
architectures generally improves the robustness of our 
normalizing flow~\citep{lakshminarayanan2016, alsing2019}.
For the bispectrum, the posteriors predicted by each individual 
flow in the ensemble are in good agreement. 

In $q_{\bphi}(\btheta\given\bfi{x})$, $\btheta$ represents the 5 cosmological and 
9 HOD parameters.
The prior of our posterior estimate is set by the parameter distribution of our training 
set.
Since the $N$-body simulations used for our forward modeled catalogs are evaluated over
a Latin-Hypercube, we use uniform priors over the cosmological parameters, 
$\{\Omega_m, \Omega_b, h, n_s, \sigma_8\}$. 
The prior ranges fully encompass the {\em Planck} priors.
For the HOD parameters, we use the same conservative priors from \citetalias{simbig_letter} 
and \citetalias{simbig_mock}.
Next, we describe our summary statistic $\bfi{x}$.

\subsection{Summary Statistic: the Galaxy Bispectrum} \label{sec:fm}
With \simbig~we can derive robust cosmological constraints using any summary 
statistic of the galaxy distribution that we can accurately forward model. 
In this work, we apply \simbig~to the first higher-order statistic: 
the galaxy  bispectrum. 
The bispectrum, $B(k_1, k_2, k_3)$, is the three-point correlation function in 
Fourier space and measures the excess probability of different triangle configurations 
$(k_1, k_2, k_3)$ over a random distribution. 
In this work, we focus solely on the monopole of the redshift-space bispectrum, 
$B_0(k_1, k_2, k_3)$.

To measure $B_0$, for both observed and forward modeled galaxy samples, we use 
the~\cite{scoccimarro2015} redshift-space bispectrum estimator, implemented in 
the $\mathtt{pySpectrum}$ python package\footnote{\url{https://github.com/changhoonhahn/pySpectrum}}.
The estimator uses Fast Fourier Transforms with grid
size $N_{\rm grid} = 360$ and box size $(1800\,h^{-1}{\rm Mpc})^3$. 
The estimator accounts for the survey geometry using a random catalog that has the same
radial and angular selection functions as the observed catalog but with a much
larger number of objects (>4,000,000).
When measuring $B_0$, we include the same \cite{feldman1994} weights as in
\citetalias{simbig_letter}.
For the observed galaxy sample, we also include angular systematic weights to 
account for stellar density and seeing conditions as well as redshift failure weights. 
We do not include weights for fiber collisions, since this effect is included in the 
\simbig~forward model.  

We measure $B_0$ in triangle configurations defined by $(k_1, k_2, k_3)$ bins of 
width $\Delta k = 0.0105\,h/{\rm Mpc}$, three times the fundamental mode
$k_f = 2\pi/(1800\,h^{-1}{\rm Mpc})$.
For $k_{\rm max} = 0.5\,h/{\rm Mpc}$, $B_0$ has 10,052 total triangle configurations. 
In practice, we use the reduced bispectrum instead of the bispectrum
to reduce the dynamic range of the summary statistic\footnote{For simplicity, we will refer use $B_0$ to refer to both the bispectrum and reduced bispectrum.}: 
\begin{align}
    &Q_0(k_1, k_2, k_3) = \nonumber\\ 
    &\frac{B_0(k_1, k_2, k_3)}{P_0(k_1)P_0(k_2) + P_0(k_1)P_0(k_3) + P_0(k_2)P_0(k_3)},
\end{align}
where $P_0(k)$ represents the monopole of the power spectrum.

We present $B_0(k_1, k_2, k_3)$ and $Q_0(k_1, k_2, k_3)$ for 200 out of 20,000 
randomly selected subset of the training set in Figure~\ref{fig:b0k}. 
We only show a subset of 1,354 triangle configurations with 
$k_1, k_2, k_3 \leq k_{\rm max} = 0.25\,h/{\rm Mpc}$ for clarity. 
We order the triangles by looping through $k_3$ in the inner most loop 
and $k_1$ in the outer most loop satisfying $k_1 \geq k_2 \geq k_3$.
For reference, we include $B_0$ of the observed CMASS sample (black) with 
uncertainties estimated using the $\mathtt{TEST0}$ simulations, which we describe
in the next section.
The $B_0$ of the training dataset has a broad range that fully encompasses the 
observed $B_0$. 

\begin{figure*}
\begin{center}
    \includegraphics[width=\textwidth]{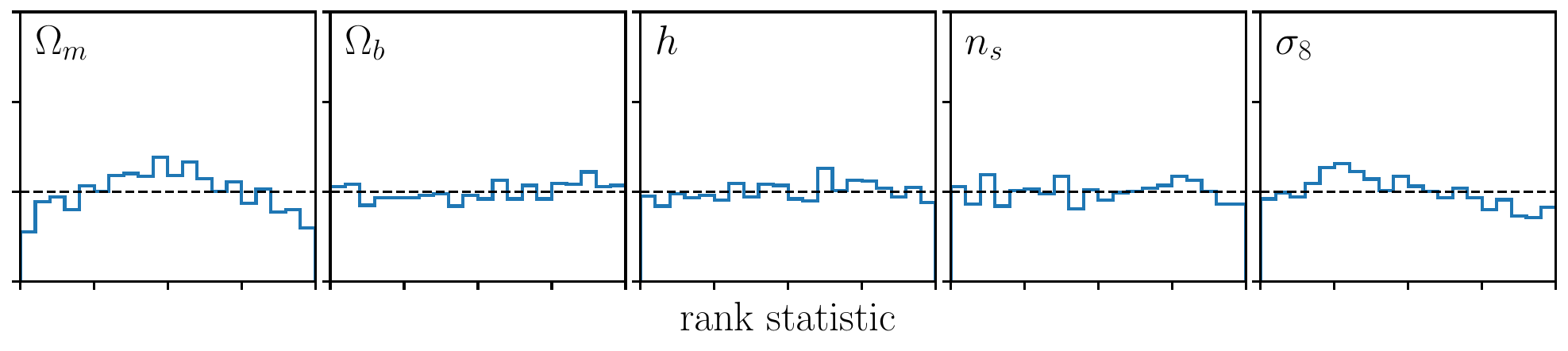}
    \caption{\label{fig:sbc}
        The NDE accuracy test that shows the SBC validation of the 
        \simbig~$B_0(k_{123} < 0.5\,h/{\rm Mpc})$ posterior estimate. 
        We present the distribution of the rank statistics, which are derived 
        by comparing the true parameter values to the inferred marginalized 1D
        posteriors. 
        The rank statistics are calculated using 2,000 validation simulations 
        that were excluded from training the posterior estimate. 
        For an accurate estimate of the true posterior, the rank statistic would 
        be uniformly distributed (black dashed). 
        Overall, we estimate unbiased posteriors of all of the
        $\Lambda$CDM cosmological parameters. 
    }
\end{center}
\end{figure*}

\subsection{Posterior Validation}  \label{sec:cross}
Before applying our \simbig~$B_0$ posterior estimator, 
$q_{\bphi}(\btheta\given\bfi{x})$, to observations, we validate
that it can robustly infer unbiased posteriors of the $\Lambda$CDM cosmological parameters. 
First, we assess whether $q_{\bphi}$ accurately estimate the posterior across the 
parameter space of the prior. 
We call this the ``NDE accuracy test''. 
In principle, with a sufficiently large training set and successful minimization, 
$q_\phi$ is guaranteed to accurately estimate the true posterior, since we train it 
by minimizing the KL divergence with the true posterior.
In our case, however, we have a limited number of simulations. 

We use the 2,000 validation simulations that were excluded from the training of
our posterior estimate (Section~\ref{sec:sbi}). 
In Figure~\ref{fig:sbc}, we present the simulation-based calibration~\citep[SBC;][]{talts2020} 
for the $\Lambda$CDM cosmological parameters. 
For each validation simulation, we apply $q_\phi$ to its
$Q_0(k_{123} < 0.5\,h/{\rm Mpc})$ measurement to infer the
posterior. 
Then for each cosmological parameter, we calculate the rank of the true parameter value
within the marginalized 1D posterior estimate. 
A uniform rank distribution indicates that we accurately estimate 
the true posterior (black dashed).
Overall, the rank distributions are close to uniform for all of the $\Lambda$CDM 
cosmological parameters. 
For $\Omega_m$ and $\sigma_8$, the distributions have a slight $\cap$-shape, which  
indicate that our $\Omega_m$ and $\sigma_8$ posterior estimates are slightly 
broader than the true posterior (\ie~underconfident). 
Since this means that our cosmological constraints will be conservative, 
we conclude that $q_{\bphi}$ is sufficiently accurate.

\begin{figure*}
\begin{center}
    \includegraphics[width=\textwidth]{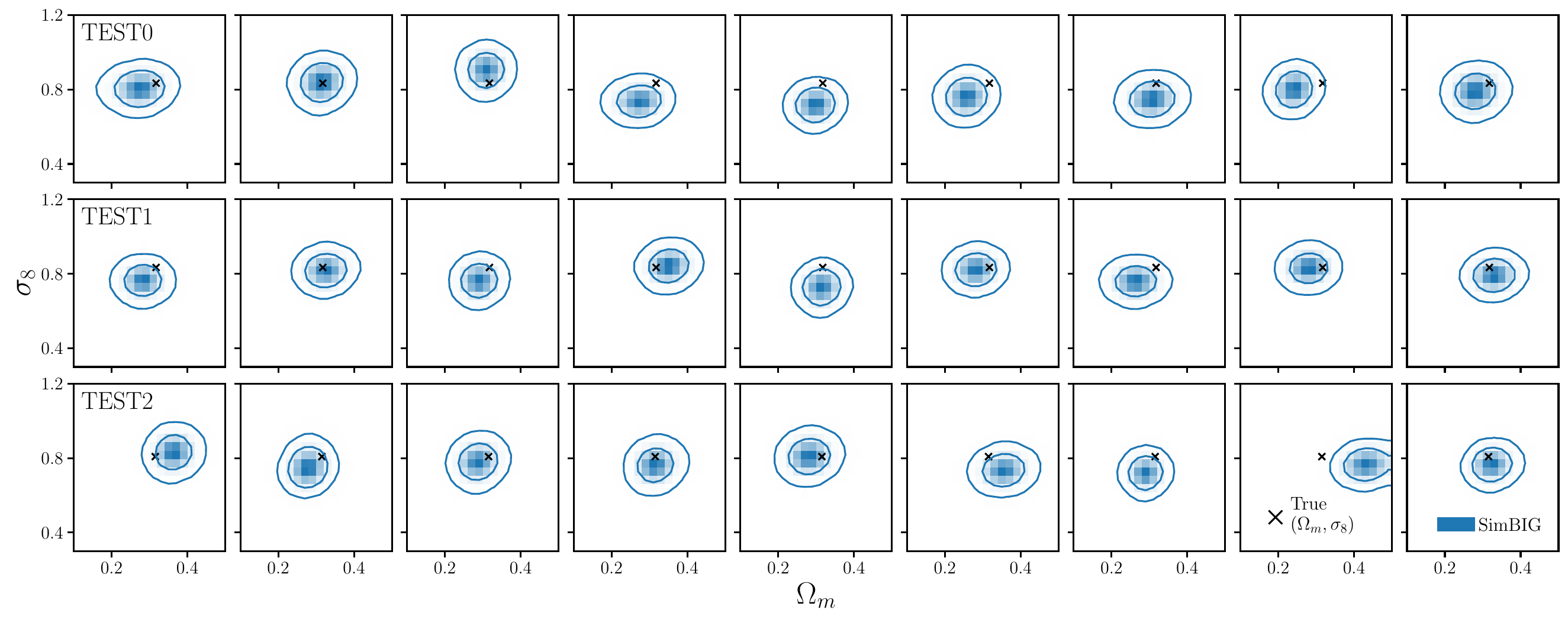}
    \caption{\label{fig:test_post}
    Posteriors of $(\Omega_m, \sigma_8)$ inferred using the \simbig~bispectrum 
    analysis for a random subset of the $\mathtt{TEST0}$ (top), $\mathtt{TEST1}$ 
    (center), and  $\mathtt{TEST2}$ (bottom) simulations.
    We mark the 68 and 84 percentiles of the posteriors with the contours. 
    We also include the true $(\Omega_m, \sigma_8)$ of the test simulations in
    each panel (black $\times$). 
    The comparison between the posteriors and the true parameter values
    qualitatively show good agreement for each test simulations. 
    }
\end{center}
\end{figure*}

Next, we verify the robustness of our $B_0$ posterior with the {\sc SimBIG} ``mock 
challenge.''
The \simbig~forward model, or {\em any} forward model, makes 
modeling choices and assumptions that, in 
detail, do not reflect the actual Universe. 
To account for this, \simbig~is designed to be highly flexible so
that we can robustly marginalize over the complex physical 
processes that govern galaxy formation and the galaxy-halo 
connection.  
Nevertheless, a summary statistic may be sensitive to the
specific choices made in the forward model.
More importantly, this can bias the inferred cosmological 
parameters. 
We, therefore, assess whether this is the case for $B_0$ 
and validate that we can derive unbiased cosmological parameter 
constraints. 

We use 2,000 test simulations in the three test sets described in
\citetalias{simbig_mock}: $\mathtt{TEST0}, \mathtt{TEST1}$, 
and $\mathtt{TEST2}$.
$\mathtt{TEST0}$ consists of 500 ``in distribution'' simulations 
built using the same forward model as the training set: 
{\sc Quijote} $N$-body, {\sc Rockstar} halo finder, and the 
full \simbig~HOD. 
$\mathtt{TEST1}$ and $\mathtt{TEST2}$ are ``out of distribution'' 
simulations. 
$\mathtt{TEST1}$ are constructed using {\sc Quijote} $N$-body,
the Friend-of-Friend halo finder~\citep[FoF;][]{davis1985}, and a 
simpler HOD model. 
Lastly, $\mathtt{TEST2}$ consists of 1,000 ``out of distribution'' 
simulations built using {\sc AbacusSummit} $N$-body 
simulations~\citep{maksimova2021}, {\sc CompaSO} halo
finder~\citep{hadzhiyska2022}, and the full \simbig~HOD.
Each test set is constructed using a different forward model. 
Hence, they serve as a stringent test sets for the robustness 
of the \simbig~$B_0$ analysis.

We run $q_{\bphi}$ on the $B_0$ of all of the test sets and 
derive a posterior for each simulation. 
In Figure~\ref{fig:test_post}, we present the 
$(\Omega_m, \sigma_8)$ 
posteriors for a randomly selected subset of the test simulations. 
We present posteriors for $\mathtt{TEST0}$, $\mathtt{TEST1}$, and
$\mathtt{TEST2}$ simulations in the top, center, and bottom panels, respectively. 
The contours represent the 68 and 95 percentiles of the posteriors. 
In each panel, we mark the true $(\Omega_m, \sigma_8)$ value of 
the test simulation (black x). 
Each test simulation is a unique realization of a CMASS-like galaxy 
catalog subject to cosmic variance. 
We, therefore, do not expect the true $(\Omega_m, \sigma_8)$ value 
to lie at the center of each of the posteriors. 
Instead, we note that for the majority of the randomly selected test
simulations, the true parameter values lie within the 68 and 95 
percentiles \simbig~posteriors. 

Next, we assess the robustness more quantitatively.
In \citetalias{simbig_mock}, we used SBC, or coverage, to assess the
robustness of the posterior estimates. 
This assessment, however, requires that the parameters of the test simulations sample 
the full prior distribution. 
Otherwise, the distribution of the rank statistic is not guaranteed to be
uniform, even for the true posterior. 
The test simulations are evaluated at fiducial values of the cosmological parameters.
Consequently, we use a different approach and assess the robustness 
by comparing the $B_0$ likelihoods of the different test sets. 
If $B_0$ is sensitive to variations in the forward model, there 
will be significant discrepancies among the likelihoods of the
test sets. 

\begin{figure*}
\begin{center}
    \includegraphics[width=\textwidth]{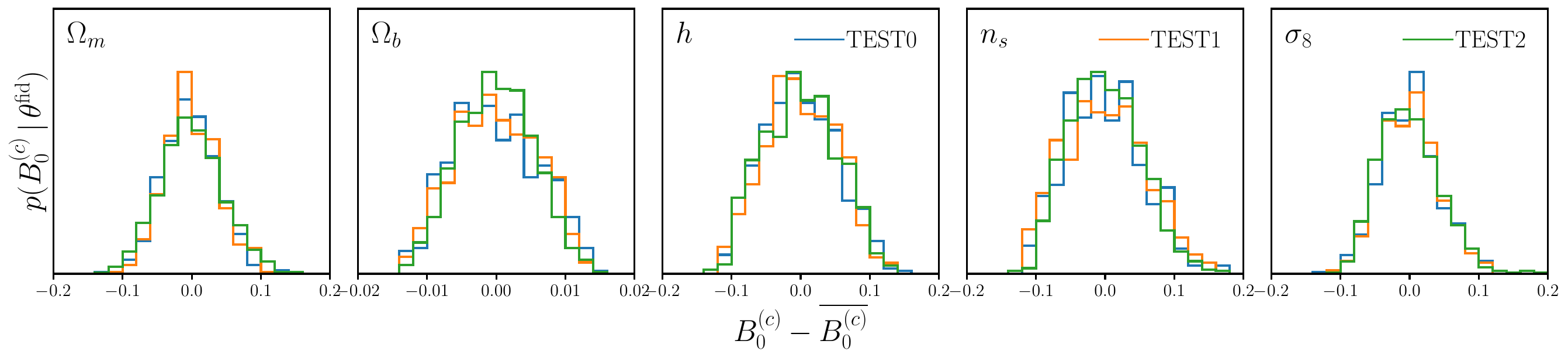}
    \caption{\label{fig:postmean}
    Comparison of the compressed bispectrum likelihood, 
    $p(B_0^{(c)}\given \theta_{\rm fid})$,  computed on the three sets of 
    test simulations: 
    $\mathtt{TEST0}$ (blue), $\mathtt{TEST1}$ (orange), and
    $\mathtt{TEST2}$ (green).
    $B_0^{(c)}$ is derived by taking the mean of the marginalized 1D 
    \simbig~$B_0(k_{123} < 0.5\,h/{\rm Mpc})$ posterior for the $\Lambda$CDM parameters, an optimal
    compression of the cosmological information in $B_0$. 
    In each panel, we mark the corresponding $\Lambda$CDM parameters.
    The likelihoods are at the fixed fiducial cosmologies and parameter
    values of the test sets. 
    We present the distribution of $B_0^{(c)} - \overline{B_0^{(c)}}$
    because $\mathtt{TEST2}$ simulations are constructed using different 
    fiducial parameter values than the $\mathtt{TEST0}$ and
    $\mathtt{TEST1}$ simulations.
    Overall, {\em we find excellent agreement among the likelihoods
    of the different test simulations and conclude that our $B_0$ 
    analysis is robust to modeling choices in our forward model}.
    }
\end{center}
\end{figure*}

In practice, comparing the $B_0$ likelihoods is challenging since 
$B_0(k_{123} < 0.5\,h/{\rm Mpc})$ is 10,052-dimensional.
We instead compare the likelihoods of the compressed $B_0$, $B_0^{(c)}$, as show in
Figure~\ref{fig:postmean} for $\mathtt{TEST0}$ (blue), $\mathtt{TEST1}$ (orange), 
and $\mathtt{TEST2}$.
For the compression, we use the mean of the marginalized 1D
\simbig~$B_0$ posterior for the $\Lambda$CDM cosmological 
parameters: 
$B_0^{(c)} = \sum_{j=1}^{N} \btheta_j/N$ where $\btheta_j \sim q_{\bphi}(\btheta\given B_0)$.
We use $N=10,000$ samples to estimate the mean. 
Each panel represents a dimension of $B_0^{(c)}$ that corresponds to 
one of the $\Lambda$CDM parameters.
This is a near-optimal compression of the cosmological information 
in $B_0$, since $q_{\bphi}$ accurately estimates the true 
posterior.

We present the distribution of $B_0^{(c)} - \overline{B_0^{(c)}}$, 
where $\overline{B_0^{(c)}}$ is the average $B_0^{(c)}$ 
instead of $B_0^{(c)}$. 
This is because the $\mathtt{TEST2}$ simulations 
are constructed using a different set of fiducial parameter 
values than the $\mathtt{TEST0}$ and $\mathtt{TEST1}$ simulations.
Overall, we find excellent agreement among the $B_0^{(c)}$ 
likelihoods with no significant discrepancies.
We also find similar levels of agreement when we use other 
summaries of the marginalized \simbig~$B_0$ posterior 
(\emph{e.g.} standard deviation, 16$^{\rm th}$ percentile) for 
the compression. 
Given the good agreement of $B_0^{(c)}$ likelihoods among the 
test sets, we conclude that our $B_0$ analysis is sufficiently
robust to the modeling choices in our forward model.

%% file: results.tex
\section{Results} \label{sec:results}

\begin{figure*}
\begin{center}
    \includegraphics[width=0.9\textwidth]{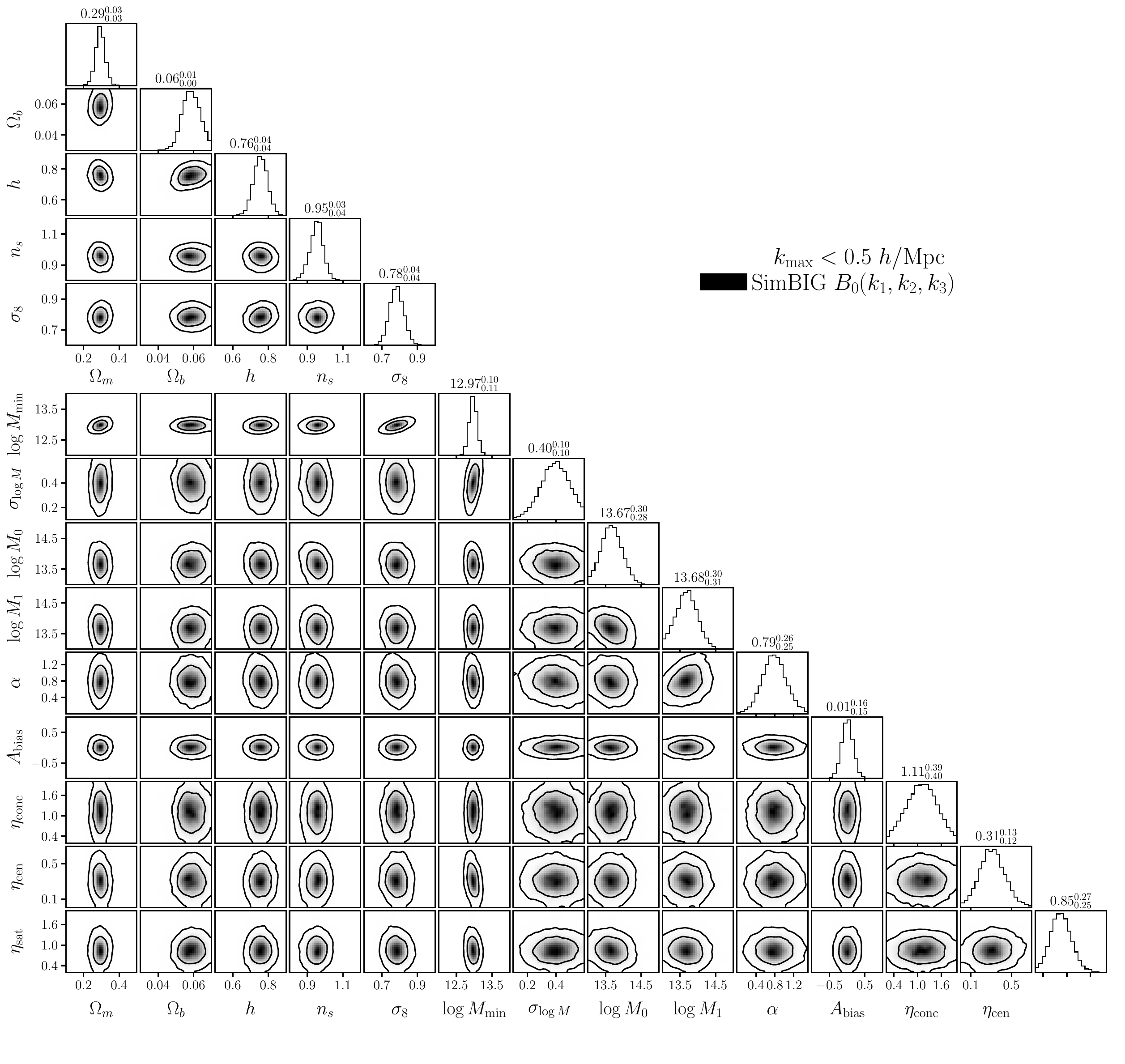}
    \caption{\label{fig:post_full}
    Posterior distribution of all parameters inferred using the 
    \simbig~$B_0$ analysis to ${k_{\rm max} < 0.5\,h/{\rm Mpc}}$ from 
    BOSS CMASS SGC.
    In the top set of panels, we present the cosmological parameters. 
    In the bottom, we present the halo occupation parameters.
    The axis ranges of the panels represent the prior range. 
    {\em 
    We place significant constraints on all $\Lambda$CDM parameters and a
    number of the halo occupation parameters} ({\em e.g.} $\log M_{\rm min}$,
    $\log M_0$, and $\eta_{\rm sat}$).
    }
\end{center}
\end{figure*}

\begin{figure*}
    \centering
    \includegraphics[width=17.8cm]{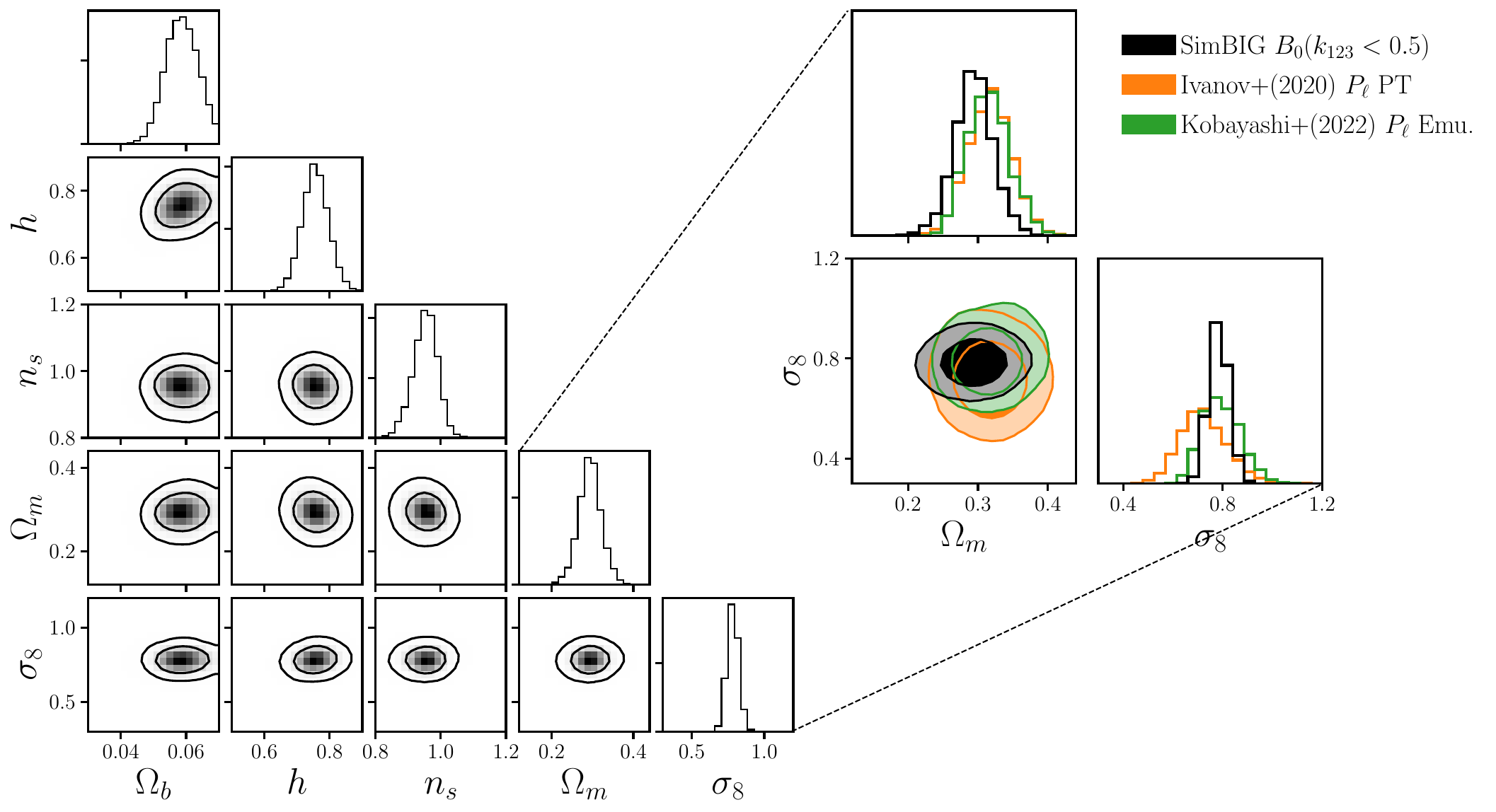}
        \caption{
        {\em Left}: 
        Posterior of cosmological parameters inferred from $B_0$ using \simbig. 
        In the diagonal panels we present the marginalized 1D posterior of each
        parameter.  
        The other panels present the 2D posteriors that illustrate the 
        degeneracies between two parameters.
        The contours mark the 68 and 95 percentiles.
        By robustly analyzing $B_0$ down to non-linear regimes, 
        $k_{\rm max} = 0.5\,h/{\rm Mpc}$, we place significant constraints on
        all $\Lambda$CDM parameters without any priors from BBN of CMB
        experiments. 
        {\em Right}: 
        We focus on the posteriors of $\Omega_m$ and $\sigma_8$), the
        parameters that can be most significantly constrained by galaxy
        clustering alone.  
        We derive
        $\Omega_m = 0.293^{+0.027}_{-0.027}$
        and 
        $\sigma_8 = 0.783^{+0.040}_{-0.038}$. 
        Our $\Omega_m$ and $\sigma_8$ constraints are >10 and 50\% tighter than 
        the $P_\ell(k < k_{\rm max} = 0.25\,h/{\rm Mpc})$ constraints from a
        PT approach~\citep[][orange]{ivanov2020} and an emulator
        approach~\citep[][green]{kobayashi2022}.
        This improvement comes from simultaneously exploiting higher-order and
        non-linear cosmological information. 
        }
    \label{fig:post}
    \centering
\end{figure*}

In Figure~\ref{fig:post_full}, we present the posterior distribution of all 
parameters inferred from the CMASS bispectrum monopole with 
$k_{\rm max} < 0.5\,h/{\rm Mpc}$ using \simbig. 
The top and bottom sets of panels present the posterior of the cosmological and
halo occupation parameters, respectively. 
The diagonal panels present the 1D marginalized posteriors; the rest of the
panels present marginalized 2D posteriors of different parameter pairs. 
The contours represent the 68 and 95 percentiles and the ranges of the panels 
match the prior.
We also list the 50, 16, and 84th percentile constraints on the parameters
above the diagonal panels. 

Focusing on the $\Lambda$CDM cosmological parameters
(Figure~\ref{fig:post}), we find that the \simbig~$B_0$ analysis
tightly constrains {\em all} of them.
This is without relying on any priors from Big Bang Nucleosynthesis 
(BBN) or cosmic microwave  background (CMB) experiments that are
typically used in galaxy clustering analyses~\citep[\emph{e.g.}][]{ivanov2020, philcox2021, kobayashi2022}.
We derive $\Omega_b= 0.059^{+0.005}_{-0.005}$, 
$h= 0.756^{+0.040}_{-0.039}$, and 
$n_s = 0.954^{+0.033}_{-0.040}$. 
For the growth of structure parameters (right panels) 
we derive: 
$\Omega_m = 0.293^{+0.027}_{-0.027}$
and $\sigma_8 = 0.783^{+0.040}_{-0.038}$. 

Our $B_0$ analysis places significantly tighter constraints than 
$P_\ell(k)$ for the same BOSS SGC sample from previous works.  
Compared to the~\citetalias{simbig_letter} \simbig~$P_\ell(k < k_{\rm max}{=}0.5)$ 
analysis, our $\Omega_m$ and $\sigma_8$ 
constraints are both 1.7$\times$ tighter. 
This $P_\ell$ analysis, however, goes beyond standard analyses and 
includes cosmological information on non-linear scales. 
If we compare to a standard PT $P_\ell(k < k_{\rm max}{=}0.25\,h/{\rm Mpc})$ 
analysis~\citep[][$\Omega_m = 0.317^{+0.031}_{-0.032}$ and 
$\sigma_8 = 0.719^{+0.100}_{-0.085}$; orange]{ivanov2020}, 
our $\Omega_m$ and $\sigma_8$ constraints are 1.2 and 2.5$\times$ tighter. 
Our constraints are also 1.1 and 2.0$\times$ tighter than the $P_\ell(k<0.25\,h/{\rm Mpc})$ constraints 
from \cite{kobayashi2022} ($\Omega_m = 0.314^{+0.031}_{-0.030}$ and
$\sigma_8 = 0.790^{+0.083}_{-0.072}$; green).
They use a theoretical model based on a halo power spectrum emulator and a halo 
occupation framework. 
These comparisons clearly illustrate that the cosmological information in 
both higher-order statistics and non-linear scales is {\em substantial}. 

\begin{figure}
\begin{center}
    \includegraphics[width=0.5\textwidth]{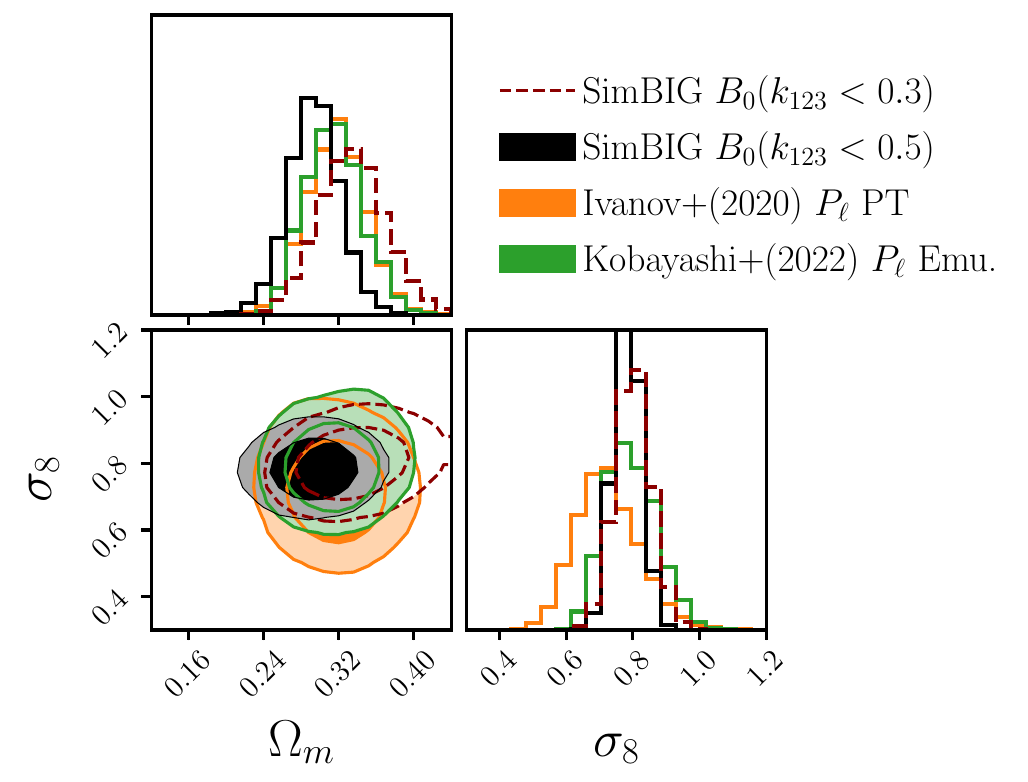}
    \caption{\label{fig:post_lowkmax}
        ($\Omega_m, \sigma_8$) posterior from the \simbig~$B_0$ 
        analysis to $k_{\rm max} = 0.3\,h/{\rm Mpc}$ (red dashed). 
        For comparison, we include posteriors from $P_\ell$ analyses
        (\citealt{ivanov2020}, orange; \citealt{kobayashi2022}, 
        green) and the \simbig~$B_0(k_{123} < 0.5\,h/{\rm Mpc})$ 
        analysis (black). 
        The contours represent the 68 and 95 percentiles.
        We find overall good agreement among the posteriors. 
        Furthermore, the improvement we find from 
        $B_0(k_{123} < 0.3\,h/{\rm Mpc})$ over $P_\ell$ is
        consistent with the improvement from $B_0$ found in the 
        literature~\citep[\emph{e.g.}][]{damico2022}.
        Our $B_0(k_{123} < 0.3\,h/{\rm Mpc})$ posterior is 
        significantly broader than our 
        $B_0(k_{123} < 0.5\,h/{\rm Mpc})$.
        This demonstrates that there is additional higher-order 
        cosmological information in the non-linear regime, 
        $0.3 < k < 0.5\,h/{\rm Mpc}$, that we can robustly 
        analyze using \simbig.
    }
\end{center}
\end{figure}

Next, we analyze $B_0$ to $k_{\rm max}=0.3\,h/{\rm Mpc}$ to examine how much 
of the improvement in our $B_0$ constraints comes from the non-linear scales alone. 
In Figure~\ref{fig:post_lowkmax}, we present the 
\simbig~$B_0(k_{123}{<}0.3\,h/{\rm Mpc})$ posterior (red dashed) 
on $\Omega_m$ and $\sigma_8$. 
We include posteriors from~\cite{ivanov2020} (orange), 
\cite{kobayashi2022} (green), and 
\simbig~$B_0(k_{123} < 0.5\,h/{\rm Mpc})$ (black).  
The contours represent the 68 and 95 percentiles of the posteriors.
We find overall good agreement among the posteriors.  
Compared to the $P_\ell$ constraints, the \simbig~$B_0(k_{123} < 0.3)$ analyses
improves $\sigma_8$ by $\sim$1.33$\times$.  
The improvement is more modest than the improvement from \simbig~$B_0(k_{123} < 0.5)$ 
and is broadly consistent with the \cite{damico2022} constraints from analyzing 
the $B_0$ to $k_{\rm max}{=}0.23\,h/{\rm Mpc}$
and bispectrum quadrupole, $B_2$, to $k_{\rm max}{=}0.08\,h/{\rm Mpc}$. 
\cite{philcox2021} and \cite{ivanov2023} recently found more modest improvements 
from the bispectrum ($\sim$1.1$\times$). 
They, however, only include the bispectrum monopole and multipoles, respectively, 
out to  $k_{\rm max}{=}0.08\,h/{\rm Mpc}$.
We refrain from a more detailed comparison since we analyze a subsample of BOSS
galaxies.
Nevertheless, the comparison illustrates that the $B_0$ on non-linear scales 
contains significant additional cosmological information. 

The \simbig~$B_0(k_{123} < 0.5)$ produces significantly tighter cosmological 
constraints than $P_\ell$ analyses because we exploit both non-Gaussian and 
non-linear cosmological information.
For $\sigma_8$, the 2$\times$ improvement in precision is roughly equivalent 
to analyzing a galaxy sample with >4$\times$ the volume using the standard
approach. 
This improvement is made possible by the \simbig~forward modeling 
approach that is not only able to accurately model galaxy clustering to 
$k_{\rm max} = 0.5\,h/{\rm Mpc}$ but also robustly account for observational
systematics. 

Interestingly, the improvements from the \simbig~$B_0$ analysis enable us to  
inform recent ``cosmic tensions'', despite only using 10\% of the full BOSS 
volume. 
These tensions refer to the discrepancies between the late time and early time 
measurements of $S_8 = \sigma_8 \sqrt{\Omega_m/0.3}$ and the Hubble constant, $H_0$,
that have been growing in statistical significance with recent 
observations~\citep[for a recent review see][]{abdalla2022}. 
They have increased the scrutiny on $\Lambda$CDM and have led to a slew of 
theoretical works to explore modifications or alternatives to 
$\Lambda$CDM~\citep[\emph{e.g.}][]{meerburg2014, chudaykin2018, divalentino2020}. 

For $S_8$, our \simbig~$B_0$ constraint $S_8 = 0.774^{+0.056}_{-0.053}$ lies slightly
above the constraints from weak lensing (WL) 
experiments~\citep[\eg~][]{asgari2021, amon2022, secco2022, dalal2023, sugiyama2023, des_kids2023}. 
We do not find significant tension with either the CMB or WL experiments. 
Our \simbig~$B_0$ analysis also places significant constraints on $H_0$, especially
when we combine our posterior with a prior on $\omega_b = \Omega_b/h^2 =0.02268\pm0.00038$ 
from BBN using importance sampling~\citep{aver2015, cooke2018, schoneberg2019}:
$H_0=67.6^{+2.2}_{-1.8}$. 
We find a lower value of $H_0$ that is in good agreement with CMB and other
galaxy clustering constraints. 

%% file: discuss.tex
\section{Discussion} \label{sec:discuss}
The \simbig~SBI approach relies on accurate forward modeling of the observed galaxy 
distributions such that the simulated and observed data are statistically indistinguishable.
To achieve this, the \simbig~forward model is designed to be highly flexible and 
mitigate the impact of model misspecification. 
It uses $N$-body simulations that can accurately model the non-linear matter 
distribution, a halo finder that robustly determines the position and velocities of 
dark matter halos, and a highly flexible state-of-the-art HOD. 

Despite these modeling choices, the \simbig~forward model does not account for all 
possible effects that may impact galaxy clustering. 
For example, it does not include the effect of baryons on the matter clustering. 
Instead, since it has a subpercent effect on the matter bispectrum at 
$k < 0.5\,h/{\rm Mpc}$~\citep[\eg~][]{foreman2019}, we rely on the HOD model 
to implicitly account for the impact.
Furthermore, we do not include redshift evolution and additional observational 
systematics (\eg~imaging incompleteness). 
We refer readers to \citetalias{simbig_mock} for a more detailed discussion on the caveats 
of our forward model.

There are also caveats to our posterior validation for $B_0$.
For instance, the comparison of the $B_0^{(c)}$ likelihoods only demonstrates the 
robustness near the fiducial cosmologies of the test simulations. 
Furthermore, some cosmological information may be lost in the $q_{\bphi}$-based 
compression scheme.
This would then potentially underestimate the discrepancies in the full $B_0$ 
likelihood.
Addressing either of these limitations, however, requires a substantially larger 
suite of simulations evaluated across the full prior space. 
We reserve developing more stringent and efficient validation of the posterior and 
summary statistic to future work. 

Significant challenges still remain when applying forward modeling approaches to 
upcoming surveys. 
They will need to be accompanied by continual improvements to the forward model 
and validation. 
There are also challenges in extending \simbig~to the large volumes and the 
different galaxy samples of upcoming surveys. 
Nevertheless, in this work we demonstrate the clear advantages of forward
modeling: by extracting cosmological information using higher-order statistics
and on non-linear scales we can {\em double} the precision of $\sigma_8$ constraints 
and significantly improve the constraints of all $\Lambda$CDM parameters. 
In the \cite{wave1}, we will present forecasts \simbig~analyses applied to upcoming galaxy 
surveys: DESI, PFS, and {\em Euclid}. 

%% file: summary.tex
\section{Summary} \label{sec:summary}
We present the \simbig~cosmological constraints from analyzing the galaxy
bispectrum monopole, $B_0(k_1, k_2, k_3)$, on non-linear scales to $k_{\rm max}=0.5\,h/{\rm Mpc}$.  
\simbig~provides a forward modeling framework that uses SBI to perform highly 
efficient cosmological inference using NDE with normalizing 
flows~(\citetalias{simbig_letter} and \citetalias{simbig_mock}). 
It enables us to leverage the predictive power of $N$-body simulations to accurately 
model higher-order clustering on small scales, which is currently inaccessible with 
standard PT analyses. 
It also allows us to more robustly include observational systematics that significantly 
impact galaxy clustering measurements. 

After validating the accuracy and robustness of our analysis using 2,000 test 
simulations constructed using three different forward models, we conduct the 
\simbig~$B_0(k_{123} < 0.5\,h/{\rm Mpc})$ analysis on a subset of CMASS 
galaxies in the SGC of SDSS-III BOSS.
We derive significant constraints on all $\Lambda$CDM parameters ($\Omega_m, \Omega_b, h, n_s, \sigma_8$) without any external priors. 
Compared to standard power spectrum analyses, we infer 1.2 and 2.4$\times$ tighter
constraints on $\Omega_m=0.293^{+0.027}_{-0.027}$ and 
$\sigma_8= 0.783^{+0.040}_{-0.038}$. 
We verify that this improvement comes from higher-order cosmological information on
non-linear scales and, when restricted to larger scales, our constraints are consistent
with previous bispectrum analyses. 

In this work, we apply \simbig to $\sim$10\% of the full BOSS volume due to the 
limited volume of our $N$-body simulations.
Despite the smaller volume, we derive growth of structure, 
$S_8 = \sigma_8 \sqrt{\Omega_m/0.3}$,
constraints competitive with other cosmological probes and BOSS analyses of the full volume. 
Our $S_8 = 0.774^{+0.056}_{-0.053}$ constraint is statistically consistent with 
both CMB and weak lensing experiments. 
We also derive a constraint on $H_0=67.6^{+2.2}_{-1.8}\,{\rm km\,s^{-1}\,Mpc^{-1}}$ 
by combining our posterior with a $\omega_b$ prior from BBN. 
Our $H_0$ constraint is consistent with early universe constraints from CMB and other 
LSS analyses. 

Even with the limited volume of our observations, we derive competitive constraints 
on $S_8$ and $H_0$ by exploiting additional cosmological information in higher-order 
clustering on non-linear scales. 
Extending \simbig~to the full BOSS volume would roughly improve the precision of
our constraints by $\sim$3$\times$.
In an accompanying paper \cite{wave1}, we will present forecasts of 
\simbig~clustering analyses of upcoming spectroscopic galaxy surveys
(\eg~DESI, PFS, {\em Euclid}) 
and demonstrate that it has to be potential to produce the leading cosmological
constraints from LSS. 
\cite{wave1} will also compare the $B_0$ constraints from this work to 
\simbig~constraints derived from field-level inference using convolutional 
neural networks~\citep{simbig_cnn} and the wavelet scatter transform~\citep{simbig_wst}.